\documentclass[11pt]{article}

\usepackage[utf8]{inputenc}
\usepackage{microtype}
\usepackage{amsmath,amsthm,amssymb, color, tikz, mathtools}
\usepackage{enumerate, enumitem} 
\usepackage{graphicx}
\usepackage{thm-restate}
\usepackage[hidelinks]{hyperref}
\usepackage{cleveref}
\usepackage{float}
\usepackage[margin = 2.50cm]{geometry}
\usepackage{comment}
\usepackage[ruled,vlined]{algorithm2e}
\usepackage{algpseudocode}
\usepackage{mathrsfs}

\renewcommand{\hat}{\widehat}

\newcommand{\R}{\mathbb{R}}
\newcommand{\E}{\mathbb{E}}

\newcommand{\cL}{\mathcal{L}}

\newtheorem{theorem}{Theorem}[section]

\newtheorem{remark}[theorem]{Remark}

\newtheorem{lemma}[theorem]{Lemma}

\newtheorem{defi}[theorem]{Definition}

\newcommand{\pmone}{\{-1, 1\}}
\newcommand{\zone}{\{0, 1\}}
\newcommand{\ftwo}{\mathbb{F}_2}
\renewcommand{\tilde}{\widetilde}
\renewcommand{\Tilde}{\widetilde}

\newcommand{\dA}{\delta_{\cL}}
\newcommand{\daffine}{\delta_{\textsf{affine}}}

\newcommand{\equ}{\mathsf{EQU}}

\newcommand{\liniso}{\mathsf{LinIso}}

\title{ 
Spectral Shadows: When Communication Complexity Meets\\
Linear Invariance Testing 
}

\author{
Swarnalipa Datta\footnote{Indian Statistical Institute, Kolkata, India}
\and
Arijit Ghosh\footnotemark[1]
\and 
Chandrima Kayal\footnote{Universit\'e Paris Cit\'e, CNRS, IRIF, Paris, France}
\and
Manaswi Paraashar\footnote{ Indian Institute of Technology, Hyderabad, India}
\and 
Manmatha Roy\footnotemark[1]
}

\date{}

\begin{document}

\maketitle

\begin{abstract}
In this short note, we initiate the study of the Linear Isomorphism Testing Problem in the setting of communication complexity, a natural linear algebraic generalization of the classical Equality problem. Given Boolean functions \( f, g : \mathbb{F}_2^n \to \{-1, +1\} \), Alice and Bob are tasked with determining whether \( f \) and \( g \) are equivalent up to a nonsingular linear transformation of the input variables, or far from being so. This problem has been extensively investigated in several models of computation, including standard algorithmic and property testing frameworks, owing to its fundamental connections with combinatorial circuit design, complexity theory, and cryptography. However, despite its broad relevance, it has remained unexplored in the context of communication complexity, a gap we address in this work.

Our main results demonstrate that the approximate spectral norm of the input functions plays a central role in governing the communication complexity of this problem. We design a simple deterministic protocol whose communication cost is polynomial in the approximate spectral norm, and complement it with nearly matching lower bounds (up to a quadratic gap). In the randomised setting with private coins, we present an even more efficient protocol, though equally simple, that achieves a quadratically improved dependence on the approximate spectral norm compared to the deterministic case, and we prove that such a dependence is essentially unavoidable. 

These results identify the approximate spectral norm as a key complexity measure for testing linear invariance in the communication complexity framework. As a core technical ingredient, we establish new junta theorems for Boolean functions with small approximate spectral norm, which may be of independent interest in Fourier analysis and learning theory. 

\end{abstract}


\section{Introduction}
\label{sec: intro}

Communication complexity, introduced by Yao~\cite{Yao79}, studies the amount of communication required between two parties, traditionally called Alice and Bob, to compute a function whose input is distributed between them. More precisely, Alice receives input $x$ from a domain $D_1$, Bob receives input $y$ from a domain $D_2$, and their goal is to compute a known function $F : D_1 \times D_2 \to \{0,1\}$ while minimizing the number of bits exchanged. This model has been extensively studied under various constraints, most notably deterministic protocols, where the parties' messages are fully determined by their inputs and communication history, and randomized protocols, which may use shared or private randomness and allow for a small probability of error, typically at most $1/3$. The distinction between deterministic and randomized communication complexity is fundamental, as randomness often enables exponential savings in communication cost.

Several classical problems have played a central role in the development of communication complexity theory. The Equality problem, where Alice and Bob must determine if their $n$-bit inputs are identical, exhibits a stark contrast between deterministic and randomized settings: while any deterministic protocol requires $\Theta(n)$ bits, randomized protocols can solve the problem using only $O(\log n)$ bits. The Set Disjointness problem is another canonical example, known for its high randomized complexity and it's important role in proving several lower bounds in different areas of theoretical computer science, e.g streaming and property testing. Other important problems include Inner Product, Gap Hamming Distance, Indexing, and Pointer Chasing, each offering unique challenges and insights. Understanding the relationships between various complexity measures, such as in the long-standing log-rank conjecture, remains one of the most intriguing open questions in the field.

Among these, the Equality problem stands out for its simplicity and foundational importance. In this setting, Alice and Bob each receive an $n$-bit string $x, y \in \{0,1\}^n$ and must decide whether $x = y$. Any deterministic protocol requires $\Omega(n)$ bits of communication, since in the worst case, Bob must verify each bit of Alice’s input. In contrast, randomized protocols, particularly those with public coins, can solve the problem using only $O(1)$ bits via simple fingerprinting techniques that detect inequality with high probability. In the private-coin model, the best known protocols require $O(\log n)$ bits~\cite{Yao79}, reflecting the cost of simulating shared randomness.

In this work, we consider a natural linear algebraic generalisation of the Equality problem, where the inputs to Alice and Bob are Boolean functions instead of bit strings, and equality is relaxed to equivalence under invertible linear transformations of the domain. Formally, given Boolean functions $f, g : \mathbb{F}_2^n \to \pmone$, we say that $f$ and $g$ are linearly isomorphic if there exists a matrix $M \in \mathrm{GL}_n(\mathbb{F}_2)$ such that
\[
g(x) = f(Mx) \quad \text{for all } x \in \mathbb{F}_2^n.
\]
Here \( \mathrm{GL}_n(\mathbb{F}_2) \) denotes the group of invertible \( n \times n \) matrices over the field \( \mathbb{F}_2 \).

Linear isomorphism of Boolean functions arises in a wide range of areas, including the design of efficient combinatorial circuits~\cite{CV03, BA04, YB12}, the construction of error-correcting codes~\cite{ND06, DT01, XH94}, and cryptographic protocols~\cite{CC07, OD12}. A notable example is provided by Reed–Muller codes, which are invariant under affine transformations of their input variables. This symmetry plays a key role in the development of fast decoding algorithms, and a detailed survey appears in~\cite{Abbe2020-aw,AbbeSSY23}. The linear isomorphism problem is also fundamental in multivariate public-key cryptography, where security often relies on the hardness of determining whether two systems of multivariate polynomials are equivalent under affine/linear transformations. Known as the isomorphism of polynomials problem, it lies at the heart of several cryptographic schemes~\cite{poly06}.

From a computational perspective, the linear isomorphism problem remains notoriously hard. Deciding whether two Boolean functions are linearly isomorphic is coNP-hard even when the functions are given in compact representations such as disjunctive normal form. The problem lies in the second level of the polynomial hierarchy $\Sigma_2^p$, but is not known to be in coNP. Agrawal and Thierauf~\cite{agariso} showed that unless the polynomial hierarchy collapses to $\Sigma_3^p$, the problem is not $\Sigma_2^p$-complete. The most efficient known algorithm, presented in~\cite{acmHypergraphIsomorphism}, reduces the task to Hypergraph Isomorphism and runs in $2^{O(n)}$ time. The linear isomorphism problem has also been explored in the property testing framework. In an important result, Wimmer and Yoshida~\cite{WY13} showed that if a Boolean function has small spectral norm, then one can efficiently test whether it is linearly isomorphic to an unknown function, or far from it using a number of queries that is polynomial in the spectral norm of the known function. Their tester only requires oracle access to the truth table of the unknown function. 

Yet, despite its foundational nature and broad relevance, the linear isomorphism problem has remained completely unexplored in the realm of communication complexity. In this work, we take the first step toward closing this gap. Our focus is on the following gap version of the linear isomorphism problem in the communication complexity setting.

\begin{defi}[The $\liniso_n$ problem]
\label{defi:linear-iso-communication-problem}
Let $\epsilon, \omega \geq 0$. Alice and Bob are each given Boolean functions  
$f, g : \ftwo^n \to \pmone$, with the promise that one of the following two cases holds:
\[
 \quad \dA(f, g) \leq \epsilon, \qquad 
 \text{or,} \qquad
 \quad \dA(f, g) \geq \epsilon + \omega,
\]
where the linear distance, $\dA(f,g)$ between $f$ and $g$ is defined as
\[
\dA(f, g) := \min_{M \in \mathrm{GL}_n(\ftwo)} \Pr_{x \in \ftwo^n} \left[ f(Mx) \neq g(x) \right].
\]
The goal is to determine which of the two cases holds; that is, to compute the function
\[
\liniso_n(f, g) :=
\begin{cases}
1 & \text{if } \dA(f, g) \leq \epsilon, \\
0 & \text{if } \dA(f, g) \geq \epsilon + \omega.
\end{cases}
\]
\end{defi}

We investigate the communication complexity of $\liniso_n$ under two standard models. In the deterministic model, Alice and Bob must decide whether $f$ and $g$ are linearly isomorphic using a deterministic protocol without any randomness, minimizing the number of bits exchanged. In the randomized model, the parties may use private randomness, and the protocol must succeed with probability at least $2/3$ on all inputs. Our goal is to understand the communication cost of solving $\liniso_n$ in these settings.

To proceed further, we introduce several foundational concepts from Fourier analysis over the Boolean cube. Fourier analysis on $\ftwo^n$ provides a natural and powerful framework for studying the structure of Boolean functions. Every function $f : \ftwo^n \to \mathbb{R}$ admits a unique representation of the form
\[
f(x) = \sum_{\alpha \in \ftwo^n} \widehat{f}(\alpha) \chi_{\alpha}(x),
\]
where the characters $\chi_{\alpha}(x) = (-1)^{\langle \alpha, x \rangle}$ are parity functions defined using the standard inner product over $\ftwo^n$. A central analytic quantity associated with a Boolean function is its spectral norm, defined by
\[
\|\widehat{f}\|_1 := \sum_{\alpha \in \ftwo^n} |\widehat{f}(\alpha)|,
\]
which plays a key role in understanding the complexity, learnability, and structural properties of Boolean functions~\cite{KushilevitzM93, TsangWXZ13, ShpilkaTV17, GirishT021}. To capture approximation in our context, we consider the approximate spectral norm, which quantifies how well a Boolean function can be approximated, pointwise, by a real-valued function with small spectral norm. For $\gamma \geq 0$, the $\gamma$-approximate spectral norm of a function $f$ is defined as
\[
\|\widehat{f}\|_{1,\gamma} := \inf \Big\{ \|\widehat{g}\|_1 : g : \ftwo^n \to \mathbb{R},\ \text{and } \forall x \in \ftwo^n,\ |f(x) - g(x)| \leq \gamma \Big\}.
\]
This quantity generalizes the spectral norm and can be significantly smaller, reflecting the compressibility of $f$ in the Fourier domain under bounded pointwise error. The approximate spectral norm has also found important applications in quantum query and communication complexity, highlighting its relevance across different computational models~\cite{TCS-107, BansalS21}.

\subsection{Our contribution} 
\label{ssec:our_contribution}

We study the communication complexity of the $\liniso_n$ problem under both deterministic and randomized protocols. We begin with the case of randomized protocols with access to public coins. When \( \epsilon = 0 \) and \( \omega > 0 \), the problem \( \liniso_n(f, g) \) admits a highly efficient public-coin randomized protocol that requires only \( O(1) \) bits of communication. The key idea is to view the truth tables of \( f \) and \( g \) as vectors in \( \pmone^{2^n} \). Alice (without interacting with Bob) computes a nonsingular matrix \( A \in \ftwo^{n \times n} \) such that the truth table of \( f \circ A \) is lexicographically minimal. Similarly, Bob independently computes a non-singular matrix \( B \) for \( g \). If \( f \) and \( g \) are linearly isomorphic, then \( f \circ A = g \circ B \), and the problem reduces to checking equality of these transformed functions. This can be done using standard public-coin randomized protocols with \( O(1) \) bits of communication (see~\cite[Chapter~3]{KushilevitzN97}). However, when \( \epsilon > 0 \), no such efficient randomized protocols are known for the tolerant version of the problem.

We now turn to the main contribution of this work: an analysis of the problem under deterministic and private-coin randomized protocols. Given functions $f, g: \ftwo^{n} \to \R$, define
\[
t := \min\left\{ \left\lceil \|\widehat{f}\|_{1,1/3} \right\rceil, \left\lceil \|\widehat{g}\|_{1,1/3} \right\rceil \right\}.
\]
Our main results, summarized below, establish nearly tight bounds on the communication complexity of this task in terms of the approximate spectral norm \( t \). In particular, we show that when \( t \) is small, a simple efficient protocols exist. However, as we show in this work, the dependence on \( t \) is unavoidable, even for randomized protocols with private coins.

\begin{theorem}[Deterministic Communication Complexity]
\label{theorem:deterministic-upper-lower}
Let \( f \) and \( g \) be Boolean functions held by Alice and Bob, respectively, and let \( \dA(f, g) \) denote their linear distance. Then:
\begin{enumerate}[label=(\alph*)]
  \item There exists a deterministic protocol that decides whether
  \[
  \dA(f, g) \leq \epsilon \quad \text{or} \quad \dA(f, g) \geq \epsilon + \omega
  \]
  using at most \( O(t^4 \log^2(1/\omega)) \) bits of communication.
  
  \item Furthermore, any deterministic protocol solving the above problem with \( \epsilon = 0 \) and \( \omega = 1/20 \) must communicate at least \( \Omega(t^2) \) bits.
\end{enumerate}
\end{theorem}

We now turn our attention to the randomized setting. We show that when parties have access to private randomness, there exists an efficient communication protocol for \( \liniso_n(f, g) \) in the non-tolerant setting. This protocol achieves a communication cost that is polynomial in the approximate spectral norm and exhibits a quadratically better dependence on this parameter compared to the deterministic case. Moreover, we prove that even in the randomized private-coin model, the dependence on the approximate spectral norm is unavoidable, unlike the case of public coin setting.

\begin{theorem}[Randomized Communication Complexity]
\label{theorem:randomized-private-coin-upper-lower}
Let \( f \) and \( g \) be Boolean functions held by Alice and Bob, respectively, and let \( \dA(f, g) \) denote their linear distance. Then:
\begin{enumerate}[label=(\alph*)]
  \item There exists a private-coin randomized protocol that decides whether
  \[
  \dA(f, g) = 0 \quad \text{or} \quad \dA(f, g) \geq \omega
  \]
  using at most \( O\left( \frac{t^2}{\omega^3} \right) \) bits of communication.
  
  \item Moreover, any private-coin randomized protocol for the above problem with  \( \omega = 1/20 \) must communicate at least \( \Omega(\log t) \) bits.
\end{enumerate}
\end{theorem}

Our results demonstrate that when the participating Boolean functions have small approximate spectral norm, the linear isomorphism problem admits efficient communication protocols. In both the deterministic and randomized (private-coin) settings, we obtain upper bounds with polynomial dependence on the approximate spectral norm, and show that this dependence is unavoidable. It remains an intriguing open question whether small approximate spectral norm characterizes the only class of functions for which this problem can be solved efficiently.

\subsection{Organization of the paper}

The remainder of this note is organized as follows.
In Section~\ref{section: Communication protocol for linear isomorphism testing}, we study the deterministic communication complexity of the linear isomorphism problem. We begin by presenting an efficient deterministic protocol whose communication cost depends polynomially on the approximate spectral norm of the input functions, and then establish a nearly matching lower bound.
In Section~\ref{section:randomized_setting}, we turn to the randomized setting. We first observe that in the public-coin model, the problem can be solved with constant communication, and then develop an efficient private-coin protocol with a quadratically improved dependence on the approximate spectral norm. We also derive a corresponding lower bound, showing that this dependence is essentially tight.
Finally, Section~\ref{section:discussion} concludes with a brief discussion of the implications of our results and several open questions for future research.

\section{Deterministic setting}
\label{section: Communication protocol for linear isomorphism testing}

Deterministic communication complexity studies the minimum number of bits two parties (Alice and Bob) need to exchange in order to compute a function whose input is distributed between them, assuming that they follow a deterministic protocol. The objective is to determine the worst-case communication cost required to compute the function exactly, without any use of randomness or allowance for error. In this section, we present a communication-efficient protocol for deciding linear isomorphism in this setting, along with a nearly matching lower bound.

Before proceeding, we introduce the following notation for the distance between two functions \( f, g : \mathbb{F}_2^{n} \to \{ -1, +1\} \): 
\[
\delta(f,g) := \Pr_{x \in \mathbb{F}_2^{n}}\!\left[f(x) \neq g(x)\right]
= \frac{\left|\left\{x \in \mathbb{F}_2^{n} : f(x) \neq g(x)\right\}\right|}{2^{n}}.
\]
We define the sign function \(\mathrm{sign}: \mathbb{R} \to \{ -1,  +1\}\) by setting \(\mathrm{sign}(x) = +1\) if \(x \ge 0\) and \(\mathrm{sign}(x) = -1\) otherwise.

\subsection{An efficient deterministic communication protocol}

We now prove the upper bound part of Theorem~\ref{theorem:deterministic-upper-lower}. As described earlier, Alice and Bob are given Boolean functions \( f, g : \mathbb{F}_2^n \to \pmone \), with the promise that either
\[
\dA(f, g) \leq \epsilon \quad \text{or} \quad \dA(f, g) \geq \epsilon + \omega.
\]
The task is to distinguish between these two cases. We present a deterministic protocol that solves this problem, described in Algorithm~\ref{algo: det communication protocol}.

We begin by proving a structural lemma for Boolean functions with small approximate spectral norm. This can be viewed as an approximate analogue of a classical result of Bruck and Smolensky~\cite{BS92} for functions admitting a small approximate spectral norm.

\begin{lemma}[Approximate Spectral Sampling]
\label{lemma:BS approximation}
Let \( f : \mathbb{F}_2^n \to \pmone \) be a Boolean function and let \( \gamma \in [0, 1) \). Then, for every \( \delta > 0 \), there exists a subset \( \mathcal{S} \subseteq \mathbb{F}_2^n \) and a Boolean function \( F : \mathbb{F}_2^n \to \pmone \) such that:
\begin{itemize}
    \item \( |\mathcal{S}| = O\left( \| \widehat{f} \|_{1, \gamma}^2 \cdot \frac{1}{(1 - \gamma)^2} \cdot \log(1/\delta) \right) \),
    \item \( F(x) = \mathrm{sign}\left( \sum_{\alpha \in \mathcal{S}} a_{\alpha} \chi_{\alpha}(x) \right) \) for some coefficients \( a_{\alpha} \in \{\pm1\} \),
    
    \item 
        $\delta(f,F) \leq \delta$.
%
\end{itemize}
\end{lemma}
In the proof of this lemma, we will use the following concentration inequality.

\begin{theorem}[Hoeffding's inequality]
\label{th:Hoeffdings inequality}
Let $X_1, X_2, \dots, X_n$ be independent random variables such that, for all $i \in [n]$, $a_i \le X_i \le b_i$.
Define $S_n := \sum_{i=1}^n X_i$ and $\mu := \E[S_n]$.
Then, for any $t > 0$, we have
\[
\Pr\!\left[\, |S_n - \mu| \ge t \,\right]
\le 2 \exp\!\left( -\frac{2t^2}{\sum_{i=1}^n (b_i - a_i)^2} \right).
\]
\end{theorem}
Our proof of the lemma follows the approach of Haviv and Regev~\cite[Lemma~3.1]{HavivR15}, who adapted the argument of Bruck and Smolensky~\cite{BS92}.

\begin{proof}[Proof of Lemma~\ref{lemma:BS approximation}]
Let \( h : \mathbb{F}_2^n \to \mathbb{R} \) be a real-valued function such that, for all $x \in \ftwo^{n}$, we have $|f(x) - h(x)| \leq \gamma$, and $\|\hat{h}\|_1 = \|f\|_{1,\gamma}$.
We express \( h \) using its Fourier expansion:
\begin{align*}
    h(x) = \sum_{\alpha \in \ftwo^{n}} \hat{h}(\alpha) \chi_{\alpha}(x)
     = \|\hat{h}\|_1 \cdot \sum_{\alpha \in \ftwo^{n}}
     \left(
     \frac{|\hat{h}(\alpha)|}{\|\hat{h}\|_1} \cdot \mathrm{sign}\left( \hat{h}(\alpha) \right) \chi_{\alpha}(x) 
     \right)
\end{align*}
This suggests a natural sampling-based approximation strategy.
Define a distribution \( \mathcal{D} \) on \( \ftwo^{n} \) as: for all $\alpha \in \ftwo^{n}$,
\[
\Pr_{\mathcal{D}}[\alpha] = \frac{|\hat{h}(\alpha)|}{\|\hat{h}\|_1}.
\]
Observe that 
\begin{align*}
    h(x) = \mathop{\mathbb{E}}_{\alpha \sim \mathcal{D}} \left[ \|\hat{h}\|_{1} \cdot
    \mathrm{sign}\left( \hat{h}(\alpha) \right) \chi_{\alpha}(x) \right]
\end{align*}
Sample \( \alpha_1, \dots, \alpha_T \sim \mathcal{D} \) independently, where
\[
T = O\left( \|\hat{h}\|_1^2 \cdot \frac{1}{{\beta}^2} \cdot \log(1/\delta) \right).
\]
Define the empirical approximator:
\[
\tilde{h}(x) = \frac{\|\hat{h}\|_1}{T} \sum_{i=1}^T \mathrm{sign}(\hat{h}(\alpha_i)) \chi_{\alpha_i}(x).
\]
By Hoeffding’s inequality (Theorem~\ref{th:Hoeffdings inequality}), this approximator satisfies:
\[
\Pr_{x \in \mathbb{F}_2^n}\left[\, |h(x) - \tilde{h}(x)| > \beta \,\right] \leq \delta.
\]
Set \( \beta := \frac{1 - \alpha}{10} \), and define
$
D := \left\{ x \in \mathbb{F}_2^n : |h(x) - \tilde{h}(x)| \leq \beta \right\},
$. Now, for all \( x \in D \), apply the triangle inequality:
\[
|f(x) - \tilde{h}(x)| \leq |f(x) - h(x)| + |h(x) - \tilde{h}(x)| \leq \beta + \gamma < 1.
\]
Since \( f(x) \in \pmone \), this implies that 
\[
\mathrm{sign}(\tilde{h}(x)) = f(x).
\]
Define the final hypothesis \( F(x) := \mathrm{sign}(\tilde{h}(x)) \). Then, for all \( x \in D \), we have \( F(x) = f(x) \), and so
\[
    \delta(f,F) = \Pr_{x \in \mathbb{F}_2^n}[f(x) \neq F(x)] \leq \delta.
\]
This concludes the proof.
\end{proof}

We now give the proof of Theorem~\ref{theorem:deterministic-upper-lower}~(a).

\begin{proof}[Proof of Theorem~\ref{theorem:deterministic-upper-lower}~(a)]
The proof of the theorem follows from the analysis of Algorithm~\ref{algo: det communication protocol}.

\paragraph{Correctness.}
We begin by recalling that the triangle inequality holds for linear distance as well. That is, for every $f, g, h : \mathbb{F}_2^n \to \{\pm 1\}$, we have:
\[
\dA(f, h) + \dA(h, g) \geq \dA(f, g).
\]

We now analyze the two possible cases:

\begin{itemize}
    \item \textbf{Case 1:} Suppose \( \dA(f, g) \leq \epsilon \). Then, by the triangle inequality for the linear distance,
    \[
    \dA(F, g) \leq \dA(f, g) + \dA(f, F) \leq \epsilon + \frac{\omega}{4}.
    \]
    
    \item \textbf{Case 2:} Suppose \( \dA(f, g) \geq \epsilon + \omega \). Then,
    \[
    \dA(F, g) \geq \dA(f, g) - \dA(f, F) \geq \epsilon + \omega - \frac{\omega}{4} = \epsilon + \frac{3\omega}{4}.
    \]
\end{itemize}

Since \( \epsilon + \frac{3\omega}{4} > \epsilon + \frac{\omega}{4} \), Bob can easily distinguish between the two cases by estimating \( \dA(F, g) \) up to additive error less than \( \omega/10 \). Thus, the protocol is correct.

\paragraph{Communication Cost.}
Communication occurs only in \textbf{Step-1} and \textbf{Step-4} of the protocol. Now we analyse them separately. 

\begin{itemize}
    \item In \textbf{Step-1}, Alice and Bob compare their respective approximate spectral norms in order to identify which of the two functions has the smaller value. This comparison requires exchanging \( O(\log t) \) bits.

    \item In \textbf{Step-4}, Alice transmits the truncated spectral representation of her function. By Lemma~\ref{lemma:BS approximation}, the support set \( \mathcal{S} \) of the truncated representation satisfies
    \[
    |\mathcal{S}| = O\left(t^2 \cdot \log(1/\omega)\right),
    \]
    and its span has dimension
    \[
    \ell = O\left(t^2 \cdot \log(1/\omega)\right).
    \]
    The initial \( \ell \) vectors spanning this space can each be represented using \( O(\ell) \) bits. Each subsequent vector in \( \mathcal{S} \) lies in the span and can be described via its coordinates in this basis using \( \ell \) bits.
    Therefore, the total communication in this step is bounded by
    \[
    O\left( \ell + |\mathcal{S}| \cdot \ell \right) = O\left( t^4 \cdot \log^2(1/\omega) \right).
    \]
\end{itemize}
Hence, the total communication cost of the protocol is 
$$
    O\left( t^4 \cdot \log^2(1/\omega) \right).
$$ 
This completes the proof. 
\end{proof}

\begin{remark}
The proof of Theorem~\ref{theorem:deterministic-upper-lower} also gives a deterministic communication protocol for deciding, whether $\daffine(f,g) \leq \epsilon$ or $\daffine(f,g) \geq \epsilon + \omega$ with same communication cost. This can be seen by a simple modification of the \textbf{Step-5} of Algorithm~\ref{algo: det communication protocol}, where Bob checks $g$ against $F \circ (M,a)$ for all $M \in \mathrm{GL}_{n}(\ftwo)$ and $a \in \ftwo^n$ (with the same choice of parameters).
\end{remark}

\begin{algorithm}
\SetAlgoLined
\caption{Deterministic Protocol for Deciding Linear Isomorphism}
\label{algo: det communication protocol}

\textbf{Input.} 
Alice is given \( f: \ftwo^n \to \{\pm 1\} \), and Bob is given \( g: \ftwo^n \to \{\pm 1\} \), 
with the promise that either \( \dA(f,g) \le \epsilon \) or \( \dA(f,g) \ge \epsilon + \omega \).\\[1mm]

\begin{description}[leftmargin=!,labelwidth=\widthof{\textbf{Step-5:}}]

\item[Step-1:]
Alice and Bob exchange bits alternately to determine whether 
\( \left\lceil \|\widehat{f}\|_{1,1/3} \right\rceil \le \left\lceil \|\widehat{g}\|_{1,1/3} \right\rceil \). 
Without loss of generality, assume 
\( \left\lceil \|f\|_{1,1/3} \right\rceil \le \left\lceil \|g\|_{1,1/3} \right\rceil \).

\item[Step-2:]
Alice constructs
\[
    F(x) := \operatorname{sign}\!\Big( \sum_{\alpha \in \mathcal{S}} a_\alpha \chi_\alpha(x) \Big),
\]
where \( |\mathcal{S}| = O(\|f\|_{1,1/3}^2 \log(1/\omega)) \) and \( a_\alpha \in \{\pm 1\} \) for all \( \alpha \in \mathcal{S} \),
such that \( \delta(f, F) \le \omega/4 \) (Lemma~\ref{lemma:BS approximation} guarantees existence).

\item[Step-3:]
Since \( \dim(\operatorname{span}(\mathcal{S})) \le |\mathcal{S}| \), Alice finds \( M \in \mathrm{GL}_n(\ftwo) \) so that
\[
    F \circ M (x) = \operatorname{sign}\Bigg( \sum_{i=1}^{\ell} a_{e_i} \chi_{e_i}(x) + \sum_{j=1}^{w} a_{W_j} \chi_{W_j}(x) \Bigg),
\]
where \( \ell + w = |\mathcal{S}| \), \( W_j \in \operatorname{span}\{e_1, \dots, e_\ell\} \), 
and \( e_i \) are standard basis vectors.

\item[Step-4:]
Alice sends Bob:
\begin{itemize}
    \item The integer \( \ell \) and coefficients \( a_{e_1}, \dots, a_{e_\ell} \) using \( O(|\mathcal{S}|) \) bits, and
    \item If \( \ell < |\mathcal{S}| \), each \( W_j \) (as an \(\ell\)-bit string) along with \( a_{W_j} \).
\end{itemize}

\item[Step-5:] 
            Bob constructs the function $F$.
            \begin{itemize}
                \item 
                    If $\dA(F,g) \leq \epsilon + \omega/4$ then the protocol accepts, and
                
                \item 
                    If $\dA(F,g) \geq \epsilon + 3\omega/4$ then protocol rejects.
            \end{itemize}

\end{description}
\end{algorithm}

\subsection{Lower bound for deterministic protocols}
\label{section: communication lower bound}
In this section, we prove the lower bound part of Theorem \ref{theorem:deterministic-upper-lower}, by reducing the string equality problem $\equ$ in communication complexity to the problem of linear isomorphism of Boolean functions $\liniso$. To describe the reduction, we first formally define $\equ$ and a nontolerant variant of $\liniso$ problem. 

\begin{defi}[$\equ_n$]
\label{defi: equality comm}
    Alice and Bob get inputs $x \in \{0,1\}^n$ and $y \in \{0,1\}^n$, respectively, and their goal is to compute $\equ_n(x,y)$, which is defined as:
        \[
        \equ_n(x,y) = 
        \begin{cases} 
        1 & \text{if } x = y \\
        0 & \text{otherwise}
        \end{cases}
        \]
    Namely, determine whether or not their inputs are equal.
\end{defi}

\begin{defi}[${\Tilde{\liniso}}_{\ell}$]
\label{defi: promised linear iso comm}
    Alice and Bob get inputs $f : \ftwo^{\ell}  \to \pmone$ and $g : \ftwo^{\ell}  \to \pmone$  respectively. Their goal is to compute ${\Tilde{\liniso}}_{\ell}(f,g)$, which is defined as:
        \[
        {\Tilde{\liniso}}_{\ell}(f,g) = 
        \begin{cases} 
        1 & \text{if } \dA(f,g) = 0 \\
        0 & \text{if } \dA(f,g) > \omega \\ 
        \end{cases}
        \]
Observe that ${\tilde{\liniso}}_{\ell}$ obtained from $\liniso_{\ell}$ by taking $\epsilon = 0$.
\end{defi}
Now we show the existence of a special kind of mapping, called $\Phi$. The main utility of this mapping lies in the fact that it kills linear relationships among input strings. Before giving the details of $\Phi$ we need to introduce some notations that will simplify our presentation. Let $T$ be a string of length $2^p$. This string can be viewed as the truth table of a Boolean function, which we will denote as $f_T : \mathbb{F}_2^p \to \mathbb{F}_2$.

\begin{lemma}[$\Phi-$Map]
\label{lemma: phi map}

    Let $n > 0$ and $\omega \in  (0, {1}/{2})$. 
    Let $m = 2^{\ell}$ such that $2cn/\omega \geq m \geq cn/\omega$ for a large enough constant $c$. Interpret each $y \in \ftwo^m$ as the truth table of the function $f_{y} : \ftwo^{\ell} \to \ftwo$. 
    Then there exists a mapping $\Phi : \mathbb{F}_2^n \to \mathbb{F}_2^{m}$ such that for all $x,y \in \ftwo^n$
    \begin{itemize}
        \item[1.] $f_{\Phi(x)} = f_{\Phi(y)}$ if $x = y$, and
        \item[2.] $\dA(f_{\Phi(x)}, f_{\Phi(y)}) \geq \omega$ if $x \neq y$.
    \end{itemize}

\end{lemma}
 First, using Lemma~\ref{lemma: phi map} we complete the proof of lower bound part of Theorem~\ref{theorem:deterministic-upper-lower}, and then later in this section, we prove Lemma~\ref{lemma: phi map}.

\subsubsection{Proof of Theorem~\ref{theorem:deterministic-upper-lower}~(b)}

Our proof relies on a class of functions obtained via a reduction from the equality function $\equ_n$. Recall that any deterministic communication protocol for $\equ_n$ must, in the worst case, exchange $\Omega(n)$ bits between Alice and Bob.

\smallskip

Consider Algorithm~\ref{algo: communication lower bound reduction}. First, observe that if $a = b$, then the corresponding functions $f$ and $g$ are linearly isomorphic, since identical functions are trivially linearly isomorphic. If $a \neq b$, then by Lemma~\ref{lemma: phi map}, we have
\[
\dA(f_{\Phi(a)}, f_{\Phi(b)}) \geq \omega.
\]

Furthermore recall that both $f_{\Phi(a)}$ and $f_{\Phi(b)}$ are defined over the domain $\mathbb{F}_2^{\ell}$, and their approximate spectral norms are bounded by $O(2^{\ell/2})$ (by Cauchy–Schwarz inequality). 

To return to our reduction, Alice and Bob now invoke an efficient deterministic protocol for the function ${\Tilde{\liniso}}_\ell$ on inputs $f_{\Phi(a)}$ and $f_{\Phi(b)}$. Suppose, for the sake of contradiction, that there exists a deterministic protocol for ${\Tilde{\liniso}}_\ell$ whose communication complexity is
\[
o\left(  \min\left\{ \left\lceil \|\widehat{f}\|_{1,1/3} \right\rceil, \left\lceil \|\widehat{g}\|_{1,1/3} \right\rceil \right\}^2 \right),
\]
where $f$ and $g$ are the functions described above (see Definition~\ref{defi: promised linear iso comm}). Then this protocol could be used to solve $\equ_n$ with only $o(n)$ bits of communication. This contradicts the known lower bound that any deterministic protocol for $\equ_n$ must exchange at least $\Omega(n)$ bits in the worst case. This concludes the proof.

\begin{algorithm}[h]
	\SetAlgoLined
	\textbf{Input.} Alice is given $a \in \ftwo^n$ and Bob is given $b \in \ftwo^n$.
        \\
        \textbf{Output.} YES if $a=b$ else NO
	\\
	\begin{description}
		\item[Step-1:]
            Alice computes $x=\Phi(a)$ and the function $f_{\Phi(a)}: 2^{\ell} \to \pmone$

        \item[Step-2:]
            Similarly, 
            Bob computes $y=\Phi(b)$ and the function $f_{\Phi(b)}: 2^{\ell} \to \pmone$
        
		\item[Step-3:]
            Alice and Bob invoke an efficient linear isomorphism protocol on $f_{\Phi(a)}$ and $f_{\Phi(b)}$
  
        \item[Step-5:] If $f$ is isomorphic to $g$,  they outputs YES, otherwise outputs NO. 
        
	\end{description}
	\caption{ From $\text{EQU}_n$ problem to $ {\Tilde{\liniso }}_\ell$}
	\label{algo: communication lower bound reduction}
\end{algorithm}

\subsubsection{Proof of Lemma~\ref{lemma: phi map}}
  \label{section: construct phi}
    
    We devise an iterative procedure for constructing the map $\Phi : \ftwo^n \to \ftwo^m$ using Algorithm \ref{construct_phi}. Recall that we assumed that $m$ is of the form $2^{\ell}$ and thus any $y \in \ftwo^m$ can be seen as the truth table of some Boolean function $f_y : \ftwo^{\ell} \to \ftwo$. 
     Two sets are maintained to facilitate the iterative procedure: 
    \begin{description}
        \item[$\mathcal{S}$:] The (partial) set of $x \in \ftwo^n$, against which the mapping has been already carried out. It is to be initialized as empty set, while at the end of procedure, it must be $\ftwo^n$. 

        \item[$\mathcal{R}$:] The set of $\Phi(x)$ where $x \in \mathcal{S}$ and all their linear isomorphisms, when interpreted as a truth table of a Boolean function on $\ell$ variables.

    \end{description}
    We also maintain the following mapping:
   \begin{description}
       \item[$\mathcal{T}$:] $\{0,1\}^m  \to \mathbf{B}_{\ell}$. A bijective mapping between binary strings of length $m = 2^\ell$ and the set of Boolean functions, denoted by $\mathbf{B}_{\ell}$, on $\ell$ variables.
   \end{description}

\begin{algorithm}
\SetAlgoLined
    \begin{description}
        \item [init:] $\mathcal{S} = \emptyset$, $\mathcal{R} = \mathbb{F}_2^m$

        \While{$\mathcal{S} \neq \mathbb{F}_2^n$}{
            \begin{enumerate}
                \item Pick some $x \in \mathbb{F}_2^n \setminus \mathcal{S}$ and set $\mathcal{S} = \mathcal{S} \cup \{x\}$
                
                \item Pick some $y$ from the set $\mathcal{R}$ and set $\Phi(x) = y$
                
                \begin{enumerate}
                    \item If no such $y$ exists, return Fail.
                \end{enumerate}

                \item Let $L(f_{\Phi(x)}, \omega) \subseteq \mathbb{F}_2^m$ be the set of truth tables of all $\ell$-bit Boolean functions\\
                that have a linear distance at most $\omega$ from $f_{\Phi(x)}$.

                \item $\mathcal{R} = \mathcal{R} \setminus L(f_{\Phi(x)}, \omega)$
                
            \end{enumerate}
        }
        return \emph{Success}.
    \end{description}
\caption{Construct $\Phi$-Map}
\label{construct_phi}
\end{algorithm}


To prove the correctness of Algorithm~\ref{construct_phi} we need the following technical lemma.

\begin{lemma}\label{lemma: number of linear iso truth table}
    Let $f: \zone^r \to \zone$ and let $L(f, \omega)$ be the set of truth tables of all $\ell$-bit Boolean functions that have a linear distance at most $\omega$ from $f_{\Phi(x)}$. Then
    $|L(f, \omega)| = O(2^{r^2} \times 2^{H(\omega) 2^r})$.
\end{lemma}
\begin{proof}
The number of non-singular linear transformations of dimension $r \times r$ is upper bounded by $2^{r^2}$. Therefore, there can be at most $2^{r^2}$ Boolean functions on $r$ variables that are linearly isomorphic to a given Boolean function $f$. Additionally, for each of these functions, the number of Boolean functions whose truth tables are at most $\omega$ fraction of entries different from it is upper bounded by $2^{H(\omega) 2^r}$, see~\cite{wiki:Hamming_bound}.

This is because, given a fixed Boolean function on $r$ variables, counting the number of neighboring functions that are at most $\omega$ fraction of entries different from it is equivalent to estimating the volume of a Hamming ball with entropy $H(\omega)$ in a field of characteristic 2 with length $2^r$. Combining these observations, we obtain the claimed bound.
\end{proof}
Now, for every iteration of the \textbf{while} loop in Algorithm~\ref{construct_phi}, from Lemma~\ref{lemma: number of linear iso truth table}, we have an upper bound of $2^{\ell^2} 2^{H(\omega)2^{\ell}}$ on the cardinality of the set $L(f_{\Phi(x)}, \omega)$ in Step-3. Since the number of iterations of the \textbf{while} loop is at most $2^n$, if one chooses $m$ in such a way that following relation holds
 \begin{align}
         2^{m} \geq 2^{n} \cdot 2^{\ell^2} 2^{H(\omega) \cdot m} \label{eq: phi correct relation gross}
\end{align}
then the algorithm never returns Fails in {\bf Step-2~(a)}, and successfully construct $\Phi$ at the end of the procedure. 

It remains to show that if $2cn/\omega \geq m \geq cn/\omega$, for a large enough constant $c$, then Equation~\eqref{eq: phi correct relation gross} holds. Observe that 
\begin{align*}
    2^{m} \geq 2^{n} \cdot 2^{\ell^2} \cdot 2^{m \cdot \mathbf{H}(\omega)}
    &\iff  m 
    \geq n + \ell^2 + {m \cdot \mathbf{H}(\omega)} \\
    &\iff  m  - {\log m}^2 - {m \cdot \mathbf{H}(\omega)} \geq n \\
    &\iff  m (1 - \frac{{\log m}^2}{m} - \mathbf{H}(\omega)) \geq n  \\
    &\iff  m (1 - \mathbf{H}(\omega) - o(1)) \geq n 
\end{align*}
Since $(1 - \mathbf{H}(\omega)) \geq \omega$ for $\omega \in (0,1/2)$, for a suitable choice of the constant $c$ we have that $m \geq cn/\omega$. Remember that for every positive integer $d$ there always exists an integer $e$ such as $d \leq 2^e \leq 2d$. We take $m$ as the smallest power of two in the interval $[cn/\omega, 2cn/\omega]$.

\section{Randomized setting}
\label{section:randomized_setting}

In randomized communication complexity, two standard models differ in how randomness is accessed: the public-coin and private-coin settings. In the public-coin model, Alice and Bob have access to a shared string of random bits that both can use throughout the protocol, allowing coordinated randomness without additional communication. In contrast, the private-coin model restricts each party to their own independently generated random bits, which are hidden from the other party.

First, we note that in the public-coin setting, the linear isomorphism problem with $\epsilon = 0$ and $\omega > 0$ can be solved with constant communication cost (see Section~\ref{ssec:our_contribution} for details). Therefore, in this section, we focus on the private-coin setting. {\bf By applying the reduction used in the previous section to prove the deterministic lower bound and using the fact that the communication complexity of the equality problem in the private-coin model is $\Omega(\log n)$, one obtains a communication lower bound of $\Omega(\log t)$.} We next show that there exists an efficient randomized communication protocol in the private-coin setting whose communication cost is polynomial in the approximate spectral norm of the input functions. This results in a communication complexity that is quadratically better in terms of spectral norm compared to the deterministic case.

We prove a key lemma that underlies our protocol. While Lemma~\ref{lemma:BS approximation} provides a function that approximates the original function with a bounded Fourier support, we strengthen this result by showing the existence of a junta that closely approximates the original function when the approximate spectral norm is bounded. This lemma will play a central role in our communication protocol in the private-coin model. Beyond this application, it may also be of independent interest in areas such as learning, property testing, and complexity theory.

\begin{lemma}[Junta approximation under linear distance]
Let \( h : \mathbb{F}_2^n \to \{\pm 1\} \) be a Boolean function whose approximate spectral norm satisfies
\(\|\widehat{h}\|_{1,1/3} \le t\), and  \( \omega \in (0,1) \).
Then there exists a Boolean function \( j : \mathbb{F}_2^n \to \pmone \) that depends on $O\left( {t^2}/{\omega^2} \right)$
coordinates (that is, a junta) and the \emph{linear distance}
satisfies $\dA(h, j) < {\omega}/{8}$.
\end{lemma}

\begin{proof}
    Let \( h^* : \mathbb{F}_2^n \to \mathbb{R} \) real-valued function 
    that pointwise \( (1/3) \)-approximates \( h \) and satisfies
\[
\|\widehat{h^*}\|_1 := \sum_{\alpha \in \mathbb{F}_2^n} |\widehat{h^*}(\alpha)| = t.
\]
Note that
\begin{align}
    \mathop{\mathbb{E}}_{x \in \ftwo^{n}}\left[ h^*(x)^{2} \right] \leq \left(1 + \frac{1}{3} \right)^2 = \frac{16}{9}. 
    \label{equation:fourier_weight_2_inequality}
\end{align}

Define the \emph{truncated Fourier approximation}:
\[
\tilde{h^*}(x) := \sum_{\substack{\alpha \in \mathbb{F}_2^n \\ |\widehat{h^*}(\alpha)| \geq \frac{\omega}{18t}}} \widehat{h^*}(\alpha) \chi_\alpha(x).
\]
Let
\[
T := \left\{ \alpha \in \mathbb{F}_2^n : |\widehat{h^*}(\alpha)| < \frac{\omega}{18t} \right\}.
\]
Using Parseval identity (see~\cite[Section~2.2]{gs001}) and linearity of Fourier transformation, we get
\begin{align*}
    \mathop{\mathbb{E}}_{x \in \ftwo^{n}} \left[ (h^*(x) - \tilde{h^*}(x))^{2} \right]
    &= \sum_{\alpha \in \ftwo^{n}} \widehat{g}(\alpha)^{2} &\mbox{where $g := h^* - \tilde{h^*}$} \\
    &= \sum_{\alpha \in \ftwo^{n}} \left( \widehat{h^{*}}(x) - \widehat{\tilde{h^{*}}}(x) \right)^{2} &\mbox{linearity of Fourier transformation} \\
    &= \sum_{\alpha \in T} \widehat{h^*}(\alpha)^2 & \\
    &< \left( \frac{\omega}{18t} \right) \cdot \sum_{\alpha \in \mathbb{F}_2^n} |\widehat{h^*}(\alpha)| &\mbox{from the definition of the set $T$} \\
    &= \left( \frac{\omega}{18t} \right) \cdot t & \\
    &= \frac{\omega}{18} &
\end{align*}


Now observe that if \( \mathrm{sign}(h^*(x)) \neq \mathrm{sign}\Big( \tilde{h^*}(x) \Big) \), then \( \big| h^*(x) - \tilde{h^*}(x) \big| \geq 2/3 \). Using this fact together with Markov's inequality gives 
\begin{align*}
    \mathop{\Pr}_{x\in \ftwo^{n}}\left[ \mathrm{sign}(h^*(x)) \neq \mathrm{sign}(\tilde{h^*}(x)) \right] 
    &\leq \mathop{\Pr}_{x\in \ftwo^{n}}\left[ 
    \big| h^*(x) - \tilde{h^*}(x) \big| \geq 2/3
    \right]  & \\
    &= \mathop{\Pr}_{x\in \ftwo^{n}}\left[ 
    \big| h^*(x) - \tilde{h^*}(x) \big|^{2} \geq 4/9
    \right]  & \\
    &\leq \left(\frac{3}{2}\right)^2 \cdot \mathop{\mathbb{E}}_{x \in \ftwo^{n}} \left[ (h^*(x) - \tilde{h^*}(x))^{2} \right] &\mbox{from Markov's inequality} \\
    &< \frac{9}{4} \cdot \frac{\omega}{18} & \\
    &= \frac{\omega}{8}. &
\end{align*}

Define the set of significant Fourier coefficients as
\[
S := \left\{ \alpha \in \mathbb{F}_2^n : |\widehat{h^*}(\alpha)| \geq \frac{\omega}{18t} \right\},
\]
and let \( k := |S| \). From Parseval’s identity (see~\cite[Section~2.2]{gs001}) and Equation~\eqref{equation:fourier_weight_2_inequality}, we have 
\[
\sum_{\alpha \in \mathbb{F}_2^n} \widehat{h^*}(\alpha)^2 = \mathbb{E}_x[h^*(x)^2] \leq \frac{16}{9}.
\]
Since each \( \widehat{h^*}(\alpha)^2 \geq \left( \frac{\omega}{18t} \right)^2 \) for all \( \alpha \in S \), it follows that
\[
k \cdot \left( \frac{\omega}{18t} \right)^2 \leq \frac{16}{9}
\quad \Rightarrow \quad
k \leq \frac{16}{9} \cdot \left( \frac{18t}{\omega} \right)^2 = \frac{5184}{9} \cdot \frac{t^2}{\omega^2} = \frac{576t^2}{\omega^2}.
\]
Therefore, 
$$
    k = O\left( {t^2}/{\omega^2} \right).
$$

Let \( \{\alpha_1, \dots, \alpha_r\} \subseteq S \) be a linearly independent subset of \( S \), where \( r := \dim(\mathrm{span}(S)) \). Define a nonsingular linear map \( A : \mathbb{F}_2^n \to \mathbb{F}_2^n \) such that \( A(\alpha_i) = e_i \) for all \( i = 1, \dots, r \), where \( \{e_i\} \) denotes the standard basis of \( \mathbb{F}_2^r \). Finally, define the Boolean function
\[
j(x) := \mathrm{sign}\left( \tilde{h^*}(Ax) \right).
\]
Observe that \( j \) depends only on \( r = O(t^2/\omega^2) \) coordinates, and satisfies $\dA(h, j) < {\omega}/{8}$.
\end{proof}

\begin{algorithm}
\SetAlgoLined
\textbf{Input.} Alice is given \( f: \mathbb{F}_2^n \to \pmone \), Bob is given \( g: \mathbb{F}_2^n \to \pmone \), with the promise that either \( \dA(f, g) = 0 \) or \( \dA(f, g) \geq \omega \).
\\

\begin{description}
    \item[Step-1:]
    Alice and Bob alternately exchange bits to determine whether 
    \[
    \left\lceil \| \widehat{f} \|_{1,1/3} \right\rceil = \left\lceil \| \widehat{g} \|_{1,1/3} \right\rceil.
    \]
    If not, they reject.

    \item[Step-2:]
    Alice computes \( f^* : \mathbb{F}_2^n \to \mathbb{R} \) that pointwise \( 1/3 \)-approximates \( f \), and satisfies \\
    \( \|f^*\|_1 = t \). She then defines the truncated Fourier approximation:
    \[
    \tilde{f^*}(x) := \sum_{\substack{\alpha \in \mathbb{F}_2^n \\ |\widehat{f^*}(\alpha)| \geq \omega/(18t)}} \widehat{f^*}(\alpha) \chi_\alpha(x).
    \]

    Let
    \[
    S := \left\{ \alpha \in \mathbb{F}_2^n : |\widehat{f^*}(\alpha)| \geq \frac{\omega}{18t} \right\},
    \]
    and let \( \{\alpha_1, \dots, \alpha_r\} \subseteq S \) be a linearly independent subset, where \( r := \dim(\mathrm{span}(S)) \).\\ 
    Define a nonsingular linear map \( A : \mathbb{F}_2^n \to \mathbb{F}_2^n \) such that
    \[
    A(\alpha_i) = e_i \quad \text{for all } i = 1, \dots, r,
    \]
    where \( \{e_i\} \) denotes the standard basis of \( \mathbb{F}_2^r \). Then define the Boolean function
    \[
    j(x) := \mathrm{sign}(\tilde{f^*}(A x)).
    \]

    Next, define a linear map \( T : \mathbb{F}_2^n \to \mathbb{F}_2^r \) such that
    \[
    T(e_i) = 
    \begin{cases}
    e_i & \text{for } 1 \leq i \leq r, \\
    0 & \text{for } r < i \leq n,
    \end{cases}
    \]
    Define the Boolean function \( p : \mathbb{F}_2^r \to \{\pm1\} \) as
    \[
    p(x) := \mathrm{sign}(j(Tx)).
    \]

    Similarly, Bob constructs a Boolean function \( q : \mathbb{F}_2^s \to \{\pm1\} \) using the same procedure\\ 
    on his function \( g \).

    \item[Step-3:]
    Alice and Bob exchange bits alternately to determine whether \( r = s \). If not, they reject.

    \item[Step-4:]
        Alice considers all functions obtained by applying linear isomorphisms to \( p \), and selects the one whose truth table is lexicographically minimal. Denote this canonical representative as \( p^* \). Similarly, Bob computes the lexicographically minimal representative \( q^* \) corresponding to \( q \).

    \item[Step-5:] 
        Alice uses her private randomness to sample a uniformly random input \( x \in \mathbb{F}_2^r \), and\\
        sends this index $x$ and $p^*(x)$ to Bob. Bob responds with the value \( q^*(x) \). If \( p^*(x) = q^*(x) \), they accept; otherwise, they reject.



\end{description}

\caption{Randomised Private Coin Protocol for Deciding Linear Isomorphism}
\label{algo: rand communication protocol}
\end{algorithm}

We now give the proof of Theorem~\ref{theorem:randomized-private-coin-upper-lower}~(a).

\begin{proof}[Proof of Theorem~\ref{theorem:randomized-private-coin-upper-lower}~(a)]
The proof of the theorem follows from the analysis of Algorithm~\ref{algo: rand communication protocol}.

\paragraph{Completeness.} Suppose \( f \) and \( g \) are linearly isomorphic Boolean functions. Then their Fourier transforms \( \widehat{f} \) and \( \widehat{g} \) are also linearly isomorphic: there exists a linear transformation \( L \in \mathrm{GL}_{n}(\ftwo) \) such that for every \( \alpha \in \mathbb{F}_2^n \),
\[
\widehat{g}(L\alpha) = \widehat{f}(\alpha).
\]
It follows that the sets of significant Fourier coefficients of \( f \) and \( g \), say \( S_f \) and \( S_g \), are related by \( S_g = L(S_f) \). Hence, the truncated approximations \( \tilde{f} \) and \( \tilde{g} \), obtained by keeping only significant coefficients, are also linearly isomorphic via \( L \). Consequently, the functions \( p = \text{sign}(\tilde{f}) \) and \( q = \text{sign}(\tilde{g}) \) are linearly isomorphic, and their canonical forms satisfy \( p^* = q^* \).

\paragraph{Soundness.} Suppose \( f \) and \( g \) are \( \omega \)-far from being linearly isomorphic, i.e.,
\[
\min_{L \in \mathrm{GL}_{n}(\ftwo)}
\ \Pr_{x \in \ftwo^{n}} \left[ f(x) \neq g(Lx) \right] \geq \omega.
\]
Let \( \tilde{f} \) and \( \tilde{g} \) be their truncated Fourier approximations, and define \( p = \text{sign}(\tilde{f}) \), \( q = \text{sign}(\tilde{g}) \). Assume \( \mathop{\Pr}_{x \in \ftwo^{n}}[f(x) \neq p(x)] \leq \omega/8 \), and similarly \( \mathop{\Pr}_{x \in \ftwo^{n}} \left[ g(x) \neq q(x) \right] \leq \omega/8 \). Then, for all linear maps \( L \), we have
\[
\Pr_{x \in \ftwo^{n}} \left[ p(x) \neq q(Lx) \right] \geq \omega - \frac{\omega}{8} - \frac{\omega}{8} = \frac{3\omega}{4}.
\]
Thus, the canonical forms \( p^* \) and \( q^* \) differ on at least a \( 3\omega/4 \) fraction of inputs.

\paragraph{Communication cost.}
To detect this difference with constant probability, it suffices to compare \( p^* \) and \( q^* \) on \( O(1/\omega) \) random inputs. In the private-coin setting, Alice chooses a random \( x \in \mathbb{F}_2^r \), where \( r = O(t^2/\omega^2) \), and sends \( (x, p^*(x)) \) to Bob. Sending \( x \) and \( p^*(x) \) costs \( O(r) \) bits. Repeating this for \( O(1/\omega) \) trials, the total communication is
$O\left( {t^2}/{\omega^3} \right)$ bits.
\end{proof}



\section{Discussion}
\label{section:discussion}

We studied the communication complexity of the linear isomorphism problem, focusing on both deterministic and private-coin randomized protocols. Our main contribution is the design of efficient protocols whose communication cost depends polynomially on the approximate spectral norm of the input functions. We also establish lower bounds, showing that a small approximate spectral norm is necessary for low-communication protocols in both settings. In the process of deriving our results, we also establish structural properties for Boolean functions with small approximate spectral norm, which may be of independent interest.

Several questions remain open. First, is the class of functions with small approximate spectral norm the only one admitting efficient communication protocols for deciding linear isomorphism? Second, can the gap between our upper and lower bounds in the private-coin randomized setting be closed? Finally, no nontrivial randomized protocol is known for the tolerant version of the problem; even partial progress in this direction would be valuable.

\section*{Acknowledgement}

Arijit Ghosh acknowledges partial support from the Science and Engineering Research Board (SERB), Government of India, through the MATRICS grant MTR/2023/001527, and from the Department of Science and Technology (DST), Government of India, through grant TPN-104427.

\bibliographystyle{alpha}

\bibliography{ref}

\end{document}